\documentstyle[12pt]{article}
\catcode`\@=11

\newcommand{\be}{\begin{equation}}
\newcommand{\ee}{\end{equation}}
\newcommand{\ba}{\begin{eqnarray}}
\newcommand{\ea}{\end{eqnarray}}
\def\NPB#1#2#3{{\it Nucl.~Phys.} {\bf{B#1}} (19#2) #3}
\def\PLB#1#2#3{{\it Phys.~Lett.} {\bf{B#1}} (19#2) #3}
\def\PRD#1#2#3{{\it Phys.~Rev.} {\bf{D#1}} (19#2) #3}
\def\PRL#1#2#3{{\it Phys.~Rev.~Lett.} {\bf{#1}} (19#2) #3}

\def\JHEP#1#2#3{{\it J. High Energy Phys.} {\bf#1} (19#2) #3}

\def\part{\partial}

\def\m\mu 
\def\n{\nu}

\def\@normalsize{\@setsize\normalsize{15pt}\xiipt\@xiipt
\abovedisplayskip 14pt plus3pt minus3pt%
\belowdisplayskip \abovedisplayskip
\abovedisplayshortskip  \z@ plus3pt%
\belowdisplayshortskip  7pt plus3.5pt minus0pt}
\def\small{\@setsize\small{13.6pt}\xipt\@xipt
\abovedisplayskip 13pt plus3pt minus3pt%
\belowdisplayskip \abovedisplayskip
\abovedisplayshortskip  \z@ plus3pt%
\belowdisplayshortskip  7pt plus3.5pt minus0pt
\def\@listi{\parsep 4.5pt plus 2pt minus 1pt
            \itemsep \parsep
            \topsep 9pt plus 3pt minus 3pt}}

\def\underline#1{\relax\ifmmode\@@underline#1\else
        $\@@underline{\hbox{#1}}$\relax\fi}
\@twosidetrue
\relax

\catcode`@=12

\catcode`\@=11

\def\section{\@startsection{section}{1}{\z@}{3.5ex plus 1ex minus
   .2ex}{2.3ex plus .2ex}{\large\bf}}

\def\thesubsection{\Roman{section}-\arabic{subsection}}

\def\ps@headings{\def\@oddfoot{}\def\@evenfoot{}
\def\@oddhead{\hbox{}\hfill
        \makebox[.5\textwidth]{\raggedright\ignorespaces --\thepage{}--
        \hfill }}
\def\@evenhead{\@oddhead}
\def\subsectionmark##1{\markboth{##1}{}} }
\renewcommand{\subsection}[1]{\addtocounter{subsection}{1}
\vspace{2.5mm}\par\noindent {\em \thesubsection . #1}\par
 \vspace{0.5mm} }

\ps@headings

\catcode`\@=12

\relax

\def\figcap{\section*{Figure Captions\markboth
        {FIGURECAPTIONS}{FIGURECAPTIONS}}\list
        {Fig. \arabic{enumi}:\hfill}{\settowidth\labelwidth{Fig. 999:}
        \leftmargin\labelwidth
        \advance\leftmargin\labelsep\usecounter{enumi}}}
 \relax
\def\tablecap{\section*{Table Captions\markboth
        {TABLECAPTIONS}{TABLECAPTIONS}}\list
        {Table \arabic{enumi}:\hfill}{\settowidth\labelwidth{Table
999:}
        \leftmargin\labelwidth
        \advance\leftmargin\labelsep\usecounter{enumi}}}
 \relax
\def\reflist{\section*{References\markboth
        {REFLIST}{REFLIST}}\list
        {[\arabic{enumi}]\hfill}{\settowidth\labelwidth{[999]}
        \leftmargin\labelwidth
        \advance\leftmargin\labelsep\usecounter{enumi}}}
 \relax

\catcode`\@=11

\def\marginnote#1{}
\newcount\hour
\newcount\minute
\newtoks\amorpm
\hour=\time\divide\hour by60
\minute=\time{\multiply\hour by60 \global\advance\minute by-
\hour}
\edef\standardtime{{\ifnum\hour<12 \global\amorpm={am}%
    \else\global\amorpm={pm}\advance\hour by-12 \fi
    \ifnum\hour=0 \hour=12 \fi
    \number\hour:\ifnum\minute<100\fi\number\minute\the\amorpm}}
\edef\militarytime{\number\hour:\ifnum\minute<100\fi\number\minute}
\def\draftlabel#1{{\@bsphack\if@filesw {\let\thepage\relax
  \xdef\@gtempa{\write\@auxout{\string
    \newlabel{#1}{{\@currentlabel}{\thepage}}}}}\@gtempa
    \if@nobreak \ifvmode\nobreak\fi\fi\fi\@esphack}
     \gdef\@eqnlabel{#1}}
\def\@eqnlabel{}
\def\@vacuum{}
\def\draftmarginnote#1{\marginpar{\raggedright\scriptsize\tt#1}}
\def\draft{\oddsidemargin -.5truein
        \def\@oddfoot{\sl preliminary draft \hfil
        \rm\thepage\hfil\sl\today\quad\militarytime}
        \let\@evenfoot\@oddfoot \overfullrule 3pt
        \let\label=\draftlabel
        \let\marginnote=\draftmarginnote
   
\def\@eqnnum{(\theequation)\rlap{\kern\marginparsep\tt\@eqnlabel}%
\global\let\@eqnlabel\@vacuum}  }
\def\preprint{\twocolumn\sloppy\flushbottom\parindent 1em
        \leftmargini 2em\leftmarginv .5em\leftmarginvi .5em
        \oddsidemargin -.5in    \evensidemargin -.5in
        \columnsep 15mm \footheight 0pt
        \textwidth 250mmin      \topmargin  -.4in
        \headheight 12pt \topskip .4in
        \textheight 175mm
        \footskip 0pt
        
\def\@oddhead{\thepage\hfil\addtocounter{page}{1}\thepage}
        \let\@evenhead\@oddhead \def\@oddfoot{} \def\@evenfoot{}  }
\def\titlepage{\@restonecolfalse\if@twocolumn\@restonecoltrue\onecolumn
     \else \newpage \fi \thispagestyle{empty}\c@page\z@
        \def\thefootnote{\fnsymbol{footnote}} }
\def\endtitlepage{\if@restonecol\twocolumn \else  \fi
        \def\thefootnote{\arabic{footnote}}
        \setcounter{footnote}{0}}  
\catcode`@=12
\relax

\def\ps@headings{\def\@oddfoot{}\def\@evenfoot{}
\def\@oddhead{\hbox{}\hfill
        \makebox[.5\textwidth]{\raggedright\ignorespaces --\thepage{}--
        \hfill }}
\def\@evenhead{\@oddhead}
\def\subsectionmark##1{\markboth{##1}{}} }

\ps@headings

\relax

\def\firstpage#1#2#3#4#5#6{
\begin{document}


\begin{titlepage}
\nopagebreak
\title{\begin{flushright}
        \vspace*{-1.8in}
        {\normalsize CPTH-S727.0799}\\[-10mm]
       {\normalsize LPT-ORSAY 99/60}\\[-10mm]
        {\normalsize ROM2F-99/23}\\[-10mm]
        {\normalsize hep-th/9908023}\\[-4mm]
\end{flushright}
\vfill {#3}}
\author{\large #4 \\[1.0cm] #5}
\maketitle
\vskip -9mm     
\nopagebreak 
\begin{abstract} {\noindent #6}
\end{abstract}
\vfill
\begin{flushleft}
\rule{16.1cm}{0.2mm}\\[-4mm]
$^{\star}${\small Research supported in part by the EEC under TMR
contract  ERBFMRX-CT96-0090.}\\[-4mm] 
$^{\dagger}${CNRS-UMR-7644.}\\[-4mm]
$^{\ddagger}${\small Laboratoire associ{\'e} au CNRS-URA-D0063.}\\
\today
\end{flushleft}
\thispagestyle{empty}
\end{titlepage}}

\date{}
\firstpage{3118}{IC/95/34} {\large\bf Brane Supersymmetry Breaking}  
{I. Antoniadis$^{\,a}$, E. Dudas$^{\,b}$ and  A. Sagnotti$^{\,a,c}$} 
{\small\sl $^a$ Centre de Physique
Th{\'e}orique$^\dagger$,  Ecole Polytechnique ,  {}F-91128
Palaiseau\\[-3mm]
\small\sl $^b$  LPT$^\ddagger$, B{\^a}t. 210, Univ. Paris-Sud, F-91405
Orsay\\[- 3mm] 
\small\sl$^{c}$ Dipartimento di Fisica, Universit{\`a} di Roma ``Tor
Vergata''\\ [-4mm]
\small\sl INFN,
 Sezione di Roma ``Tor Vergata''\\[-4mm]\small\sl Via della Ricerca
Scientifica 1, 00133 Roma, Italy}  
{We show how to construct chiral
tachyon-free perturbative orientifold models, where supersymmetry is
broken at the string scale on a collection of branes while, to lowest
order, the bulk and the other branes are supersymmetric. In higher
orders, supersymmetry breaking is mediated to the remaining sectors,
but is suppressed by the size of the transverse space or by the
distance from the brane where supersymmetry breaking primarily
occurred. This setting is of interest for orbifold models with
discrete torsion, and is of direct relevance for low-scale string
models. It can guarantee the stability of the gauge hierarchy against
gravitational radiative corrections, allowing an almost exact
supergravity a millimeter away from a non-supersymmetric world.}

The breaking of supersymmetry in String Theory is a long-standing
fundamental problem with many ramifications. It is related to the
selection of the correct vacuum state, to the cosmological constant
problem, to the lifting of flat directions for string moduli, and is a
necessary ingredient of a realistic string phenomenology.
Unfortunately, despite the recent progress in the understanding of
non-perturbative phenomena based on string dualities, little was done
on the problem of supersymmetry breaking. 

In perturbation theory, closed string vacua with spontaneously 
broken\footnote{In the sense that it can be restored by tuning a
continuous parameter.} supersymmetry can be constructed generalizing
the Scherk-Schwarz mechanism, and in particular resorting to freely
acting orbifolds \cite{ss,a}. The breaking scale is then fixed by a
compactification radius,  and realistic scenarios ask for radii of the
order of a few TeV \cite{a}.
This approach is therefore likely to be relevant if the
string scale is far below the Planck mass \cite{w}, and possibly close
to electro-weak energies
\cite{add,aadd,low}. A natural framework for such models is the type-I
string theory, where gauge interactions are localized on D-branes while
gravity propagates in the bulk \cite{aadd}.

Scherk-Schwarz compactifications were recently extended to  type I
string models in \cite{ads}, where a new  feature was pointed out: the
massless spectra of D-branes orthogonal to the coordinate used for the
breaking remain supersymmetric at the tree level.  As a result, in this
case the scale of supersymmetry breaking in  the observable world is
not directly proportional to the compactification scale, and can have
lower values, making this class of constructions more flexible and 
potentially relevant even if the string scale is moved to intermediate
values \cite{aq,interm}.

The main problem with this mechanism is the cosmological constant. The
reason is that the bulk energy density behaves generically as
$\rho_{\rm bulk}\sim 1/R^{4+d_\perp}$, where $R$ is the radius of the
coordinate used to break supersymmetry and $d_\perp$ is the
dimensionality of the space transverse to our brane world, assumed
large with respect to the string scale. The projection of this
cosmological constant on the brane is enhanced by the volume of the
transverse space $r^{d_\perp}$, and is far above ${\cal O}$(TeV$^4$).
In fact, the radius $R$ of a longitudinal direction can not be far
from the (TeV) type I string scale
$M_I$ in a perturbative setting, and as a result the brane energy
density acquires a quadratic sensitivity to the four-dimensional Planck
mass,
$\rho_{\rm brane}\sim r^{d_\perp}M_I^{4+d_\perp}\sim M_{Pl}^2M_I^2$
\cite{adpq}. On the other hand, if $R$ is transverse ($R\sim r<<{\cal
O}({\rm TeV}^{-1}$)), one obtains $\rho_{\rm brane}\sim 1/r^4$. In
both cases, the energy density on the brane is far above the TeV
scale, and this destabilizes the hierarchy that one tries to enforce.
One way out is to resort to special models with broken supersymmetry
but with a vanishing or exponentially small cosmological constant
\cite{ks}. 

Alternatively, one could conceive a different scenario, with
supersymmetry broken primordially on our brane world and a string
scale of a few TeV.  In this case the brane cosmological constant
would be, by construction, ${\cal O}(M_I^4)$, while the bulk, only
affected by gravitationally suppressed radiative corrections, would be
almost supersymmetric \cite{aadd}. In particular, one would expect
that the gravitino mass  and the other soft masses in the bulk be 
extremely small $O(M_I^2/M_{Pl}) \sim 10^{-4} $ eV for $M_I
\sim 1$ TeV.  Such small masses for scalar moduli and gauge fields
might also induce deviations from Newtonian gravity in the
(sub)millimeter region that can be experimentally tested
\cite{forces,aadd}. Moreover, the cosmological constant induced in the
bulk would be $\rho_{bulk} \sim M_I^4 /r^{d_\perp} \sim
M_I^{6+d_\perp}/M_P^2$, {\it i.e.} of order (10 MeV)$^6$ for
$d_\perp=2$. Alternatively,  brane supersymmetry breaking could  also
be of interest in models with an intermediate string scale
$\sim 10^{11}$ GeV \cite{interm}, if it occurs on a brane distant from
our world and is therefore  mediated to us by gravitational
interactions.

The purpose of this letter is to show that it is possible to construct
perturbative orientifold models \cite{carg} where supersymmetry
breaking originates from a collection of branes, while both the bulk
spectrum and the spectrum of other branes are supersymmetric at
tree-level. Whereas models with  bulk supersymmetry can naturally be
constructed in the effective field theory, for instance appealing to
non-perturbative super-Yang-Mills dynamics, we believe that a direct
string construction is of some interest, in particular to attain a
better comparison with field theory supersymmetry breaking mechanisms.

Brane supersymmetry breaking can also be induced turning on internal
magnetic fields \cite{ba}. The mechanism we are proposing shares some
properties with this setting, although supersymmetry is broken at the
string scale, but appears to avoid some of its problems, namely the
presence of tachyons and the generic  lack of gaugino masses for the
unbroken gauge group.

The rules for constructing perturbative type-I orientifolds
\cite{open,gp} rest on the modular invariance of the closed string
spectrum and on some conditions linking its Klein-bottle projection to
the open and unoriented sector.  These are to be supplemented by
Ramond-Ramond (RR)  tadpole conditions, that are directly related to
anomaly cancellations and may be regarded as global neutrality
conditions for RR charges in a compact internal space \cite{pc}.
Tadpole cancellations result from opposite contributions of boundaries
and crosscaps or, equivalently, of branes and orientifolds. 

It is possible to construct models where some of the orientifold
contributions are inverted, so that the necessary RR cancellations
require some care. A simple example of this phenomenon is afforded by
the $T^4/Z_2$ orientifold where, compatibly with the ``crosscap 
constraint'' \cite{fps}, all twisted  Klein-bottle contributions are
reversed.  These exotic Klein-bottle projections are quite
interesting, and have already led to tachyon-free non-supersymmetric
open-string  vacua \cite{nontach}. In our case, the resulting
unoriented closed  spectrum is supersymmetric and contains, aside from
the $(1,0)$ gravitational multiplet, 17 tensor multiplets and 4
hypermultiplets, but its RR tadpoles can not be canceled in the usual
way. A more sophisticated, but physically very interesting, set of
examples, is provided by the open descendants of $Z_2 \times Z_2$
models, and  in particular of those with discrete torsion, where some
of the orientifold charges are necessarily  reversed. In all these
cases, the tadpole conditions may be solved unpairing NS and R
contributions, and thus inducing brane supersymmetry breaking.  In this
letter, for the sake of brevity, we confine our attention to the
$T^4/Z_2$ case, where one obtains a chiral 6D spectrum that is free of
tachyons and satisfies all usual anomaly cancellation constraints. In
this model, that contains 32 D9 and 32 anti-D5 branes,  the
absence of tachyons can also be understood as in recent studies
\cite{sen} of stable non-BPS states of the type IIB string.

The same mechanism can be applied to the $Z_2 \times  Z_2$ models, that
will be discussed elsewhere, and in principle should offer the
possibility to deal with other orientifold models where tadpole
conditions do not admit naive supersymmetric solutions.

\section*{A six-dimensional example}

We now present an explicit 6D model containing D9 and D${\bar 5}$
(anti-D5) branes, where at tree level supersymmetry breaking is
induced on the 
${\bar 5}{\bar 5}$ and ${\bar 5} 9$ states, while the $99$ states and
the bulk (closed) spectrum are supersymmetric. The starting point in
this construction is a modification of the
$\Omega$ projection in the twisted  sector of the $T^4/Z_2$ model. This
inverts the charge of the O5 planes, and is actually  compatible with
the perturbative rules of orientifold models, but the cancellation of
RR tadpoles requires 32 D${\bar 5}$ branes.  Omitting for brevity the
contributions of the transverse bosons, the torus partition function is
\ba  {\cal T} = \frac{1}{2} |Q_o + Q_v|^2 \Lambda + 
\frac{1}{2} |Q_o - Q_v|^2 {\biggl|\frac{2 \eta}{\theta_2}\biggr|}^4 
\nonumber +
\frac{1}{2} |Q_s + Q_c|^2 {\biggl|\frac{2 \eta}{\theta_4}\biggr|}^4 +
\frac{1}{2} |Q_s - Q_c|^2 {\biggl|\frac{2 \eta}{\theta_3}\biggr|}^4 \ ,
\label{a1}
\ea  where $\Lambda$ denotes the compactification lattice. In
${\cal T}$, we have introduced the convenient combinations of  $SO(4)$
characters
\ba  Q_o &=& V_4 O_4 - C_4 C_4 \quad , \qquad Q_v = O_4 V_4 - S_4 S_4
\ , 
\nonumber \\ Q_s &=& O_4 C_4 - S_4 O_4 \quad , \qquad Q_c = V_4 S_4 -
C_4 V_4 \ , 
\label{a2}
\ea 
defined as
\be  O_4 = { \theta_3^2 + \theta_4^2 \over 2 \eta^2} \quad , \quad V_4
= { \theta_3^2 - \theta_4^4 \over 2 \eta^2} \quad , \quad S_4 =
{\theta_2^2 - \theta_1^2 \over 2 \eta^2} \quad , \quad C_4 = {
\theta_2^2 + \theta_1^2 \over 2 \eta^2} \ . \label{a3}
\ee  

Turning to the Klein bottle, let us consider the two inequivalent
choices\footnote{The world-sheet moduli for the various amplitudes,
implicit in the following, are defined as in \cite{ads}.}
\be {\cal K} = \frac{1}{4} \biggl\{ ( Q_o + Q_v ) ( P + W ) + 2
\epsilon
\times 16 ( Q_s + Q_c ){\biggl(\frac{\eta}{\theta_4}\biggr)}^2 \biggr\}
\ , 
\label{a4}
\ee  where $P$ ($W$) denotes the momentum (winding) lattice sum and
$\epsilon = \pm 1$. For both choices of
$\epsilon$, the closed string spectrum has $(1,0)$ supersymmetry, but
the two resulting projections are quite different.  The usual choice
($\epsilon=1$) leaves 1 gravitational multiplet, 1 tensor multiplet
and 20 hypermultiplets, while
$\epsilon=-1$ leaves 1 gravitational multiplet, 17 tensor  multiplets
and 4 hypermultiplets\footnote{A similar projection with a non-zero
$B_{ab}$ \cite{open} would result in 13 or 11 tensor multiplets
\cite{carlo}.}. The projected closed spectrum coincides with the one
considered in ref. \cite{bz}, but our open spectrum is markedly
different. The transverse-channel Klein bottle amplitude reads
\be
\tilde{\cal K} = \frac{2^5}{4} \biggl\{ ( Q_o + Q_v ) \biggl( v W^e  +
\frac{P^e}{v} \biggr)  + 2 \epsilon  ( Q_o - Q_v ) {\biggl(\frac{2
\eta}{\theta_2}\biggr)}^2
\biggr\} \ , \label{a5}
\ee  where  $P^e$ ($W^e$) denotes the lattice of even momenta
(windings) and
$v$ is the volume of the compact space.  The reversal of the O5 charge
respects the positivity structure at the origin of the $T^4$ lattice,
and indeed  the coefficients combine into perfect squares:
\be
\tilde{\cal K}_0 = \frac{2^5}{4} \biggl\{ Q_o \biggl( \sqrt{v}  +
\frac{\epsilon}{\sqrt{v}}\biggr)^2 + Q_v \biggl( \sqrt{v}  -
\frac{\epsilon}{\sqrt{v}}\biggr)^2 \biggr\} \ . \label{a6}
\ee  
The case $\epsilon=1$ leads to the familiar $U(16) \times U(16)$
model \cite{open,gp}, and therefore from now on we restrict our
attention to the other choice. 

RR tadpole cancellations and the modifications of
${\cal K}$ are compatible with the positivity of the open spectrum in
${\cal A}$, provided one introduces D${\bar 5}$ branes, rather than
the usual D5 branes. In the transverse channel, this choice inverts
the signs of all RR contributions in the Neumann-Dirichlet ($9 {\bar
5}$) part, but leaves all other untwisted terms unchanged. Introducing
suitable Chan-Paton charges $N,D$ and their $Z_2$ orbifold breakings
$R_N,R_D$, the transverse annulus amplitude reads
\ba
\tilde{\cal A} &=& \frac{2^{-5}}{4} \biggl\{ (Q_o + Q_v) \biggl( N^2 v
W  +
\frac{D^2 P}{v} \biggr) + 2 N D (Q'_o - Q'_v) {\biggl(\frac{2
\eta}{\theta_2}\biggr)}^2  \label{a7} \\ &+&  16 (Q_s + Q_c) \biggl(
R_N^2 + R_D^2 \biggr){\biggl(\frac{
\eta}{\theta_2}\biggr)}^2 + 8 R_N R_D ( V_4 S_4 - O_4 C_4 - S_4 O_4 +
C_4 V_4 ){\biggl(\frac{
\eta}{\theta_3}\biggr)}^2
\biggr\} \ , \nonumber
\ea  where we have also introduced primed characters, related by a
chirality change $S_4 \leftrightarrow C_4$ to the unprimed ones in
(\ref{a2}):
\ba Q'_o &=& V_4 O_4 - S_4 S_4 \quad , \qquad Q'_v = O_4 V_4 - C_4 C_4
\quad ,
\nonumber \\ Q'_s &=& O_4 S_4 - C_4 O_4 \quad , \qquad Q'_c = V_4 C_4 -
S_4 V_4 \quad . 
\label{a8} 
\ea  As usual, at the origin of the lattice the different terms
organize into perfect squares:
\ba &&\tilde{\cal A}_0 = \frac{2^{-5}}{4} \biggl\{ Q'_o \biggl( N
\sqrt{v}  +
\frac{D}{\sqrt{v}}\biggr)^2 + Q'_v \biggl( N\sqrt{v}  -
\frac{D}{\sqrt{v}}\biggr)^2  \label{a9} \\ &&\!\!+ ( V_4 S_4 - S_4 O_4
)\
\biggl( 15 R_N^2 \!+\! (R_N \!+\! 4 R_D)^2 \biggr) \!+\! ( O_4 C_4
\!-\!
 C_4 V_4 ) \biggl( 15 R_N^2 \!+\! (R_N \!-\! 4 R_D)^2 \biggr)
 \biggr\} \ . \nonumber
\ea  Here, for simplicity, all D${\bar 5}$ branes, whose geometry is
neatly displayed by the breaking terms, have been placed at the origin
of the compact space. The direct-channel annulus is obtained by an
S-transformation, and reads
\ba {\cal A} &=& \frac{1}{4} \biggl\{ (Q_o + Q_v) ( N^2 P  + D^2 W ) + 
2 N D (Q'_s + Q'_c) {\biggl(\frac{\eta}{\theta_4}\biggr)}^2 
\label{a10}
\\ &+& (R_N^2 + R_D^2) (Q_o - Q_v) {\biggl(\frac{2
\eta}{\theta_2}\biggr)}^2 + 2 R_N R_D ( - O_4 S_4 - C_4 O_4 + V_4 C_4 +
S_4 V_4 ){\biggl(\frac{
\eta}{\theta_3}\biggr)}^2 \biggr\} \ . \nonumber
\ea  

Finally, the M{\"o}bius amplitude at the origin of the lattices 
(for the definition of the hatted characters, see \cite{ads}), 
\ba
\tilde{\cal M}_0 &=& - \frac{1}{2} \biggl\{ \hat{V}_4
\hat{O}_4 \biggl( \sqrt{v}  -
\frac{1}{\sqrt{v}}\biggr) \biggl( N \sqrt{v}  +
\frac{D}{\sqrt{v}}\biggr) + \hat{O}_4
\hat{V}_4 \biggl( \sqrt{v}  +
\frac{1}{\sqrt{v}}\biggr) \biggl( N \sqrt{v}  -
\frac{D}{\sqrt{v}}\biggr) \nonumber \\ &-& \hat{C}_4 \hat{C}_4 \biggl(
\sqrt{v}  -
\frac{1}{\sqrt{v}}\biggr) \biggl( N \sqrt{v}  -
\frac{D}{\sqrt{v}}\biggr) -
\hat{S}_4 \hat{S}_4 \biggl( \sqrt{v}  +
\frac{1}{\sqrt{v}}\biggr) \biggl( N \sqrt{v}  +
\frac{D}{\sqrt{v}}\biggr)
\biggr\} \ , \label{a11}
\ea  is easily obtained combining
${\tilde {\cal K}_0}$ and ${\tilde {\cal A}_0}$ , and  allows one to
reconstruct the full transverse M{\"o}bius amplitude
\ba
 &\tilde{\cal M}& = - \frac{1}{2} \biggl\{ N v W^e ( \hat{V}_4
\hat{O}_4  + \hat{O}_4 \hat{V}_4  - \hat{C}_4 \hat{C}_4 - \hat{S}_4
\hat{S}_4 ) - \frac{ D P^e} {v} ( \hat{V}_4
\hat{O}_4  + \hat{O}_4 \hat{V}_4  + \hat{C}_4 \hat{C}_4 + \hat{S}_4
\hat{S}_4 ) \nonumber \\ &-&\!\!\!\!\! N( \hat{V}_4
\hat{O}_4 \!-\! \hat{O}_4 \hat{V}_4 \!-\! \hat{C}_4 \hat{C}_4
\!+\! \hat{S}_4 \hat{S}_4 )\left(
{2{\hat{\eta}}\over{\hat{\theta}}_2}\right)^2 
\!\!+\! D( \hat{V}_4
\hat{O}_4 \!-\! \hat{O}_4 \hat{V}_4 \!+\! \hat{C}_4 \hat{C}_4
\!-\! \hat{S}_4 \hat{S}_4)\left(
{2{\hat{\eta}}\over{\hat{\theta}}_2}\right)^2 
\biggr\} \ . \label{a12}
\ea  

Eq. (\ref{a12}) shows some marked differences with respect to the
more familiar model with $\epsilon=1$. These may be given a neat
physical interpretation, since $\tilde{\cal M}$ describes the
propagation between holes and crosscaps, or equivalently between branes
and  orientifold planes. Therefore, one can see that all D9-O9
terms, the D${\bar 5}$-O5 terms in the R-R sector and the  D${\bar
5}$-O9 terms in the NS-NS sector are as in the standard $T^4/Z_2$
orientifold, while the signs of all D9-O5 terms, of the  D${\bar
5}$-O5  terms in the NS-NS sector and of the  D${\bar 5}$-O9 terms
in the R-R sector are inverted. In particular, this implies that the
M{\"o}bius amplitude  breaks supersymmetry at tree level in the
D${\bar 5}$ sector, an effect felt by all open-strings ending on the
D${\bar 5}$ branes.

Finally, a P transformation determines the direct (open string)
amplitude
\ba  &{\cal M}& = - \frac{1}{4} \biggl\{ N P ( \hat{O}_4
\hat{V}_4  + \hat{V}_4 \hat{O}_4  - \hat{S}_4 \hat{S}_4 - \hat{C}_4
\hat{C}_4 ) -  D W ( \hat{O}_4
\hat{V}_4  + \hat{V}_4 \hat{O}_4  + \hat{S}_4 \hat{S}_4 + \hat{C}_4
\hat{C}_4 ) \nonumber \\ &-&\!\!\!\!\! N( 
\hat{O}_4 \hat{V}_4 \!-\! \hat{V}_4 \hat{O}_4 \!-\! \hat{S}_4 \hat{S}_4
\!+\! \hat{C}_4 \hat{C}_4 )\left(
{2{\hat{\eta}}\over{\hat{\theta}}_2}\right)^2  \!\!+\! D( \hat{O}_4
\hat{V}_4 \!-\! \hat{V}_4 \hat{O}_4 \!+\! \hat{S}_4 \hat{S}_4
\!-\! \hat{C}_4 \hat{C}_4)\left(
{2{\hat{\eta}}\over{\hat{\theta}}_2}\right)^2  \biggr\} \ .
\label{a13} 
\ea  Parametrizing the Chan-Paton charges as
\be  N=n_1+ n_2 \quad , \qquad D=d_1+ d_2 \quad , \qquad R_N=n_1- n_2
\quad , \qquad R_D=d_1- d_2 \ , \label{a14}
\ee  the RR tadpole conditions $N=D=32,R_N=R_D=0$
($n_1=n_2=d_1=d_2=16$)  determine the gauge group $[ SO(16) \times
SO(16) ]_9 \times  [ USp(16)
\times USp(16) ]_5$.

The massless matter representations may be read from:
\ba  &&{\cal A}_0 + {\cal M}_0 = \frac{n_1(n_1-1) + n_2(n_2-1) +
d_1(d_1+1) +  d_2(d_2+1) }{2} \ V_4 O_4 \nonumber \\
 &-& \frac{n_1(n_1-1) + n_2(n_2-1) + d_1(d_1-1) + d_2(d_2-1) }{2} \ C_4
C_4 \label{a15} \\ &+& (n_1 n_2 + d_1 d_2 ) ( O_4 V_4 - S_4 S_4 ) + (
n_1 d_2 + n_2 d_1 ) \ O_4 S_4 - (n_1 d_1 + n_2 d_2 ) \ C_4 O_4 \ . 
 \nonumber
\ea  The $99$ spectrum is supersymmetric, and comprises the (1,0)
vector multiplets for the $SO(16) \times SO(16)$ gauge group and a
hypermultiplet in the representations ${\bf\! (16,16,1,1)}$ of the
gauge group. On the other hand, the ${\bar 5} {\bar 5}$ DD spectrum is
not supersymmetric, and contains, aside from the gauge vectors of $[
USp(16)
\times USp(16) ]$, quartets of scalars in the ${\bf (1,1,16,16)}$,
right-handed Weyl fermions in the $ {\bf (1,1,120,1)}$ and in the ${\bf
(1,1,1,120)}$, and left-handed Weyl fermions in the
${\bf (1,1,16,16)}$. Finally, the ND sector is also non supersymmetric,
and comprises doublets of scalars in the ${\bf (16,1,1,16)}$ and in
the 
${\bf (1,16,16,1)}$, together with additional (symplectic)
Majorana-Weyl fermions in the ${\bf (16,1,16,1)}$ and ${\bf
(1,16,1,16)}$. These Majorana-Weyl fermions are a peculiar feature of
six-dimensional space time,  where the fundamental Weyl fermion, a
pseudoreal spinor of
$SU^*(4)$, can be subjected to an additional Majorana condition,  if
this is supplemented by a conjugation in a pseudoreal representation
\cite{wsi}. In this case, this is indeed possible, since the ND
fermions are valued in the fundamental representation of $USp(16)$.
This doubling is also a useful technical trick in six-dimensional
supergravity, where Weyl fermions are often extended to $Sp(2)$
Majorana-Weyl doublets in the
$(1,0)$ case and to $Sp(4)$  Majorana-Weyl quartets in the $(2,0)$
case.

It should be appreciated that, from the $D9$ brane point of view, 
the diagonal combination of the two $USp(16)_{\bar 5}$ gauge groups
acts as a global symmetry. This corresponds
to having complex scalars and symplectic Majorana-Weyl fermions in the
representations $16\times[{\bf (16,1)+(1,16)}]$ of the D9 gauge group.
As a result, the non-superymmetric ND spectrum looks effectively 
supersymmetric, and indeed all $9 \bar{5}$ terms do not
contribute to the vacuum energy. However, as in the 
6D temperature breaking discussed in \cite{ads},
the chirality of the fermions in $Q'_s$ is not the one
required by 6D supersymmetry.   This chirality flip 
is a peculiar feature of six-dimensional models, and does not persist
in the reduction to four dimensions
\footnote{In view of the present results, the 6D model presented in the
first reference in \cite{ads} actually contains, in addition to the
$D9$ branes, one set of 16 $D5$ branes and one set of 16 $D{\bar 5}$
branes. This justifies both the chirality changes in the twisted
spectrum of the 
$D{\bar 5}$ branes and the presence of the tachyonic mode in the
$5{\bar 5}$ annulus amplitude.}. Thus, in four dimensions
supersymmetry  would be strictly unbroken on the
$D9$ branes before turning on $D{\bar 5}$ gauge interactions.
This setting shows some marked differences with respect to
the mechanism of bulk
supersymmetry breaking (or brane supersymmetry), where the
massless ND sector was supersymmetric from both the N and D 
viewpoints \cite{ads}.

It is easy to check that all the irreducible gauge and
gravitational anomalies cancel in this model as a result of the
tadpole conditions for the RR fields. The residual anomaly polynomial
does not factorize, and reveals the need for a generalized
Green-Schwarz mechanism 
\cite{ggs}, with couplings of a more general type than those found in
supersymmetric models:
\ba A &=& \frac{1}{64}  (F_1^2 + F_2^2 + F_3^2 + F_4^2 - 2R^2)^2
 + \frac{3}{64} (F_1^2 - F_2^2 - F_3^2 + F_4^2)^2 \nonumber \\ &+&
\frac{5}{64} (F_1^2 - F_2^2 + F_3^2 - F_4^2)^2  -\frac{1}{64} (F_1^2 +
F_2^2 - F_3^2 - F_4^2)^2 \quad .
\label{a16}
\ea

Brane-antibrane interactions have been discussed recently in the
literature in the context of stable non-BPS states
\cite{sen}. Our results for the D9-D${\bar 5}$ system, restricted to
the open oriented sector, provide particular examples of type-I vacua
including stable non-BPS states with vanishing interaction energy for
all radii, as can be seen from the vanishing of the ND
annulus amplitude.

Comparing the unoriented closed and open string amplitudes (\ref{a4}),
(\ref{a10}) and (\ref{a13}) with those of  the supersymmetric $T^4/Z_2$
orientifold, it is easy to see that the former can be obtained from the
latter by a
$\pi$-rotation of the D5 branes together with a reversal of the
charge of the O5 planes. This phenomenon presents some analogies
with the deformations induced by an internal magnetic field \cite{ft}
felt only by the D5 branes. This would shift the  oscillator mode
numbers of the 55 strings by an amount
$\epsilon=\epsilon_I+\epsilon_J$, where
\be
\epsilon_I={1\over\pi}\arctan (\pi q_IH) \quad ,
\label{H}
\ee  with $q_I$ the charges of the corresponding Chan-Paton states. In
a similar fashion, it would shift the mode numbers of the 95 strings by
$\epsilon=\epsilon_I$, and the M{\"o}bius contributions by
$\epsilon=2\epsilon_I$. These oscillator shifts translate into the mass
shifts $\delta M^2=(2n+1)|\epsilon|+2\epsilon S_{int}$, where the
integer $n$ denotes the Landau level and $S_{int}$ denotes  the
internal spin. As a result, in a generic magnetic field  the internal
components of the higher-dimensional gauge fields become tachyonic.
This is not the only problem presented by type-I strings in magnetic
fields: RR tadpoles are generally non-vanishing while, from a more
phenomenological  perspective, the masses of all the gauginos of the
unbroken gauge group, that are neutral under $Q$, vanish.

Our mechanism consists partly of introducing a discrete value
$\epsilon_I=1$, outside the physical range of eq.(\ref{H}). As a
result, in the annulus the 99 and 55 contributions are unaffected,
while in the 95 terms the internal characters are interchanged,
according to
$S_4\leftrightarrow C_4$ and $O_4\leftrightarrow V_4$. On the other
hand, as we have seen,  the modifications of the M{\"o}bius amplitude
are more subtle, due to the simultaneous sign change of the O5
charge, that results into a symplectic gauge group in the 55 sector
with no massless adjoint fermions. Thus, the spectrum is tachyon free,
all RR tadpoles are canceled and there are no massless gauginos.

The breaking of supersymmetry gives rise to a vacuum energy localized
on the D$\bar{5}$ branes, and thus to a tree-level potential for the N
S moduli, that can be extracted from the corresponding uncancelled NS
tadpoles in eqs. (\ref{a6}), (\ref{a9}) and (\ref{a11}). A simple
inspection shows that the only non-vanishing ones correspond to the NS
characters
$V_4O_4$ and $O_4V_4$ associated to the 6D dilaton $\phi_6$ and to the
internal volume $v$:
\be  {2^{-5}\over 4}\left\{ \left( (N-32){\sqrt v}+{D+32 \over{\sqrt
v}}\right)^2 V_4O_4 +\left( (N-32){\sqrt v}-{D+32 \over{\sqrt
v}}\right)^2 O_4V_4
\right\} \, . \label{a17}
\ee  Using factorization and the values $N=D=32$ needed to cancel the
RR tadpoles, the potential (in the string frame) is:
\be  V_{\rm eff}=c{e^{-\phi_6}\over{\sqrt v}}=ce^{-\phi_{10}} ={c\over
g_{\rm YM}^2}\, , \label{a18}
\ee  where $\phi_{10}$ is the 10D dilaton, that determines the
Yang-Mills coupling
$g_{\rm YM}$ on the D5 branes, and $c$ is some {\it positive}
numerical  constant. The potential (\ref{a18}) is clearly localized on
the D5 branes, and is positive. This can be understood noticing that
the O9  plane contribution to vacuum energy, identified from
(\ref{a17}), is negative  and exactly cancels for $N=32$. This fixes
the D5 brane contribution to the vacuum energy, that is thus 
positive, consistently with the interpretation of this mechanism as
global supersymmetry breaking. The potential (\ref{a18}) has the usual
runaway behavior, as expected by general arguments.

An interesting application of the above mechanism is to the $Z_2 \times
Z_2$ orbifold model with discrete torsion, where nontrivial two-form
fluxes at the orbifold fixed points \cite{vw} invert the sign of the
lattice independent terms in the torus amplitude.  It is interesting
to notice that the corresponding supersymmetric spectra,
which contain chiral fermions, are inconsistent because of uncancelled
RR tadpoles. The solution to the RR tadpoles can again be
obtained introducing anti-D5 branes, along the lines described
previously. We have worked out in detail the partition functions and
the spectra of the open descendants, that will be presented
elsewhere. Here we conclude by describing some qualitative features of
the resulting models. The O$5_i$ plane charges, manifest in the Klein
bottle amplitude, are equal to
$-32 \epsilon_i$, where the signs $ \epsilon_i =\pm 1$ are restricted
by
$\epsilon_1 \epsilon_2 \epsilon_3=-1$. There are thus four independent 
possibilities, $(\epsilon_1, \epsilon_2, \epsilon_3) = (-1,-1,-1),
(-1,1,1),(1,-1,1)$ and $(1,1,-1)$. As in the 6D model discussed in the
present paper, every O$5_i$ charge flip $\epsilon_i=-1$ will ask for a
set of 32  D${\bar 5}_i$ branes. Moreover, in this case the D${\bar
5}_i$ gauge group becomes  unitary, whereas for
$\epsilon_i=1$ it is symplectic as in the supersymmetric model without
discrete torsion. However, the gauginos of unitary groups are
massless, and can acquire masses only through quantum corrections.
Indeed, in our 6D example the gauginos were massive because the
M{\"o}bius amplitude had different projections for bosons and
fermions. On the other hand, the adjoint representation for unitary
groups is not affected by the M{\"o}bius 
amplitude, and therefore the corresponding gauginos stay massless. 
The D9 gauge group changes too: it is $SO(8)_9^4$ for
 $(\epsilon_1, \epsilon_2, \epsilon_3) = (-1,-1,-1)$ and $U(8)_9^2$ for
the other 3 choices. Supersymmetry is broken at the string scale in the
sectors $9{\bar 5}_i$ and in the Dirichlet part of  the M{\"o}bius
amplitude, in analogy with the 6D model described here.  Moreover, if
$\epsilon_i=1$ and  $\epsilon_j=-1$ supersymmetry  is also broken in
the sector $5_i {\bar 5}_j$, where a tachyon appears, in
agreement with the analysis of non-BPS states of \cite{sen}. 
Finally, the
remaining NS-NS tadpoles give rise to a scalar potential localized on
the  anti-D5 branes, as in eq. (\ref{a18}).
 
It would be interesting to investigate along the same lines  other
models, as the 4d $Z_4$ orientifold,  where the tadpole conditions
have no naive solution.
 
\vskip 16pt
\begin{flushleft} {\large \bf Acknowledgments}
\end{flushleft} 

We are grateful to C. Angelantonj, G. D'Appollonio and K. Ray for
several useful discussions. A.S. would like to thank CNRS for financial
support and the Centre de Physique Theorique of the Ecole Polytechnique
for the kind hospitality extended to him during the course of this
research.

\end{document}